# FALCON: a concept to extend adaptive optics corrections to cosmological fields

F. Hammer[a], M. Puech[a], F. Assemat[a], E. Gendron[a], F. Sayède[a], P. Laporte[a], M. Marteaud[a], A. Liotard[b], F. Zamkotsian[b]

[a] Observatoire de Paris - Meudon, 5 place Jules Janssen, 92195 Meudon, France
[b] Laboratoire d'Astrophysique de Marseille, 2 place Leverrier, 13248 Marseille Cedex 4, France

**ABSTRACT**

FALCON[1] is an original concept for a next generation spectrograph at ESO VLT or at future ELTs. It is a spectrograph including multiple small integral field units (IFUs) which can be deployed within a large field of view such as that of VLT/GIRAFFE. In FALCON, each IFU features an adaptive optics correction using off-axis natural reference stars in order to combine, in the 0.8-1.8 µm wavelength range, spatial and spectral resolutions (0.1-0.15 arcsec and R=10000+/-5000). These conditions are ideally suited for distant galaxy studies, which should be done within fields of view larger than the galaxy clustering scales (4-9 Mpc), i.e. foV > 100 arcmin$^2$. Instead of compensating the whole field, the adaptive correction will be performed locally on each IFU. This implies to use small miniaturized devices both for adaptive optics correction and wavefront sensing. Applications to high latitude fields imply to use atmospheric tomography because the stars required for wavefront sensing will be in most of the cases far outside the isoplanatic patch.

## 1. DATING THE EPOCHS OF GALAXY FORMATION: DYNAMICS, CHEMISTRY AND STELLAR POPULATIONS

Galaxy formation is a complex phenomenon which extends over most of the Hubble time. The Hubble morphological classification is no longer adapted to describe galaxies when the Universe was only half its present age (z=0.7-1): galaxy morphologies were irregular and chaotic. At z > 0.5-1 star formation density was dominated by luminous infrared galaxies, which are mostly enshrouded star forming sources. At these moderate redshifts, galaxies are enough bright to allow detailed studies of their chemistry and dynamics, with proper estimates of the dust extinction. It is probable that soon, an alternative classification sequence of z~1 galaxies will emerge, relating them to those of the local Hubble sequence. Galaxy physics studies are very demanding for accurate measurements, including dynamics (Tully Fischer and fundamental planes), extinctions, star formation rates, gas abundances and stellar population syntheses. Spectral resolution in excess of ~ 2000 [2] and 3D spectroscopy[3] are pre-requisites for these studies.

At higher redshifts (z=2 to 6), most galaxies have been identified using the Lyman break drop out method. They are about 5 times less numerous than z<1 galaxy population and they are strongly clustered with a correlation length of $r_0$= 4 Mpc[4]. These numbers might be only preliminary since they are limited by the present depth of the observations to R~ 26. We presently ignore the extinction properties of Lyman beak galaxies (LBGs), and determining their (probably low) O/H abundances is almost beyond the reach of 8 meter telescopes[5]. A population of dust enshrouded starbursts have been discovered at 0.85mm by SCUBA, which is probably similar to some ultra luminous IR galaxies today. These galaxies contribute to few 10% of the sub-mm cosmic background, and are so red (and/or redshifted) that they are often not detected by 8 meter telescopes. At the highest redshifts (z>6), only few objects have been tentatively identified, generally on the basis of their single Lyα emission line. A considerable effort has been devoted to these searches, since it is predicted (and confirmed from two z~6 QSO spectra[6]) that the reoinisation epoch was occurring at such redshifts.

We believe that the next step towards a description of galaxy formation over the Hubble time requires 3D spectroscopy in near IR at spectral resolutions from 2000 to 15000. Major goal is to follow up the population of z >> 1 galaxies and to describe the main physical processes which relate them to the z=1 and then to the local galactic population (Figure 1). A similar goal, although much more modest, is underway for z < 1 galaxies, thanks to the recent implementation of FLAMES/GIRAFFE at VLT. Using this instrument, one can derive velocity fields of several z~ 1 galaxies, using the 15 Integral Fields Units (IFU) at resolution of 4500-9000 [7]. Extending these to large redshifts requires to shift to the near IR (to catch redshifted emission lines from [OII]3727 to Hα) as well as to substantially improve the spatial resolution. Another important requirement is the field of view which should be significantly larger than the correlation scales to avoid strong systematics related to the cosmic variance. These considerations are at the basis of the FALCON concept as described in Hammer et al (2001)[8].

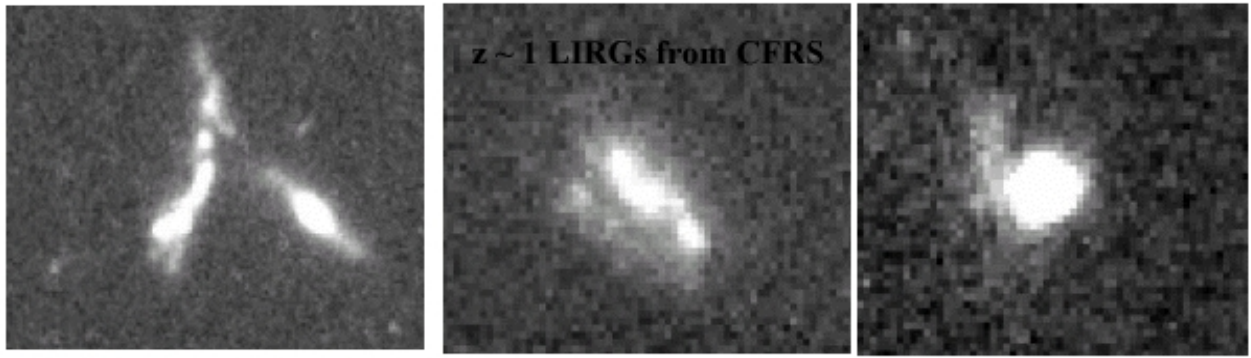

**Figure 1: examples of luminous IR galaxies at z~ 1 (field size is 50x40 kpc, from Zheng et al, 2003 [9])**

## 2. FIELD OF VIEW AND IMAGE QUALITY

Previous experience on deep fields indicates that cosmic variance can severely affect our view of the distant Universe. Major differences in number counts and of galaxy appearances have been found between the two small (WFPC2 fields, ~ 6 arcmin$^2$) Hubble Deep Fields[10]. This is illustrated by simulations of the cosmic web based on the ΛCDM theory (Figure 2). The field size should encompass the clustering scales (4 to 9 Mpc) at all redshifts, giving a minimal field of view of 100 arcmin$^2$. Most important programs at VLT are aiming at studying specific sources (LBGs, LIRGs, sub-mm ULIRGs, ellipticals, disks, etc…), for which the redshift range is limited by the spectral range of the spectrograph, leading to surface density of few 0.01 to few 0.1 per arcmin$^2$. The FALCON field of view (Φ=25 arcmin at VLT Nasmyth focus) is ideally suited for distant galaxy studies.

Several efforts have been based in trying to correct fields of few arcmin$^2$ using adaptive optics. The related scientific goals are mostly to derive accurate morphologies of distant galaxies (GEMINI MCAO system). On the other hand, ground based multi-object 3D spectrographs at moderate resolution require very large CCD formats. We believe that the optimal way to use them is to sample several individual areas of interests provided by the scientific targets. Instead of correcting the whole field, the AO correction will be performed locally on each galaxy.

Distant galaxies (z=0.5-6) are small (few arcsec$^2$) and are low surface brightness sources. It would be essential to concentrate the light within a given aperture to improve the S/N (less sky, more object) and to provide enough spatial resolution to sample their internal kinematics, chemistry and dust properties. Low surface brightness of distant galaxies naturally put limits on the spatial resolution sampling for a given collecting area (because of the spatial SNR). Ideally, we could reach a spatial resolution of FWHM ~ 0.2 arcsec (0.06 arcsec) allowing to resolve areas of 1.5 kpc at z~ 1.5 (400 pc at z~3-6) , for an 8 meter telescope (and for a 30 meter telescope, respectively).

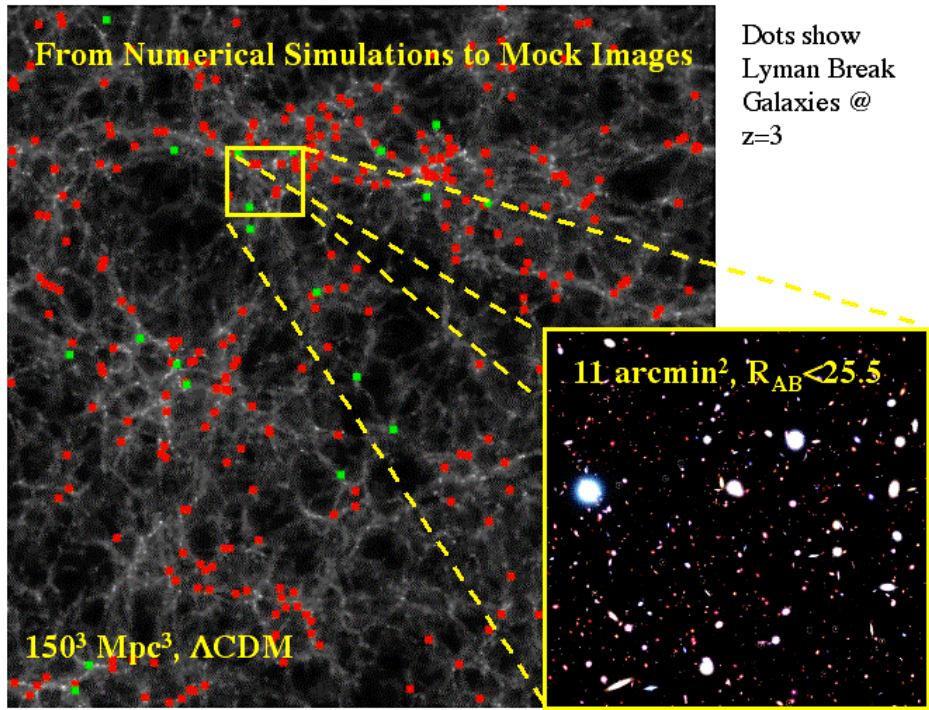

**Figure 2:** simulation from GALICs[11] of the cosmic web at z~3. Red and green points shows that LGBs are correlated with the filaments. Small fields of view would lead to strong cosmic variance when studying physical properties of distant galaxies.

## 3. FALCON ADAPTIVE OPTICS SYSTEM

In the case of astronomical AO, it is required to have a bright star (R< 16) within the isoplanatic field of radius $\theta_0$[12] to sense the wavefront. The main goal of FALCON being extragalactic astronomy, it will observe distant galaxies in directions far away from the galactic plane to avoid contamination of light by our galaxy. As these distant galaxies are very faint objects they cannot be directly used to perform wavefront sensing and unfortunately and at high galactic latitude, the surface density of stars decreases dramatically : the probability to find a suitable star for wavefront-sensing can be as low as 1%. As an example, at the north galactic pole, there are only ~ 0.1 stars with R < 16 per arcmin$^2$ [13]. This makes the sky coverage so low that classical AO is unsuitable to satisfy the scientific goals.

Laser Guide Star (LGS) have been proposed[14] to improve the sky coverage[15] of AO systems, but LGS suffer from the cone effect[16] and the tilt determination problem[17]. Applying LGS to the FALCON case (few 10 IFUs spread over 25 arcmin) requires numerous LGS (roughly one LGS per IFU) and as many tilt-stabilization systems using Natural Guide Stars (NGS). This creates technological difficulties and increases by far the cost of such an instrument.

A solution to this problem could be to use Multi-Conjugated Adaptive Optics[18,19,20,21,22] (MCAO). The goal of MCAO is to correct anisoplanatism by measuring the wavefront in several directions, reconstruct the phase perturbation in 3 dimensions and then to use several deformable mirrors (DM) conjugated to different turbulent layers located in altitude to have a good correction in an extended field of view (FOV). Thus MCAO performes corrections in the volume of turbulence, in contrary of classical AO which performes an integrated correction in a pupil plane. That is why MCAO can widen the corrected FOV since it takes into account volumic effects and thus, angular effects through this volume. The problem is that even with MCAO, it is not possible to correct a field as wide as several 100 arcmin$^2$ as it requires huge DM with a too high number of actuators. As an example, the MCAO system for the Gemini South telescope[23] will correct the turbulent phase in a 4 arcmin$^2$ wide extended FOV.

The approach we propose on FALCON differs totally from classical AO or MCAO. Instead of correcting the 25 arcmin Nasmyth FOV of the VLT as a whole, we only correct the regions of interest, i.e. the corresponding IFU superimposed to the observed galaxies[24]. To do this, we use several independent AO systems (one system per IFU).

First, as the IFU size (~2x3 arcsec$^2$) is smaller than the isoplanatic patch, a single DM conjugated to the pupil can be used to correct the wavefront. In fact, we plan to use an "hybrid" corrector componed of a adaptive lens (for tip-tilt and defocus modes correction) and a micro-DM (for higher order modes), as DM are usually unable to correct low order modes (see section 5). Both the lens and the micro-DM will be placed in pupil planes and thus no MCAO-type corrections will be performed. Second, we assume that such a corrector and its pupil relay optics can be miniaturized to be integrated into the spectroscopic IFU, i.e. the so-called adaptive button. We think this technological challenge should be achievable within a few years thanks to the development of micro-DM. Third, we assume that the wave front sensor (WFS) can be miniaturized too and fit into a so-called WFS-button, which is similar to a spectroscopic IFU, but which is located on a guide star (no spectroscopy is performed here, just wavefront sensing). Therefore, all these new components (adaptive buttons and WFS-buttons) can be handled by an IFU-positioner (which might be OzPoz at VLT). However, this architecture departs from any other usual closed-loop AO system, as there is no optical feedback from the micro-DM to WFS. One solution is to integrate a micro-DM in each WFS-button and apply to it a command identical to that of the adaptive button. This "pseudo closed loop" is based on the assumption that all the micro-DM have the same behavior, which can be a critical point and requires accurate calibration: the loop is closed by "electro-mechanical analogy". The FALCON AO loop represents therefore a difficulty and requires further studies on which we are currently working on.

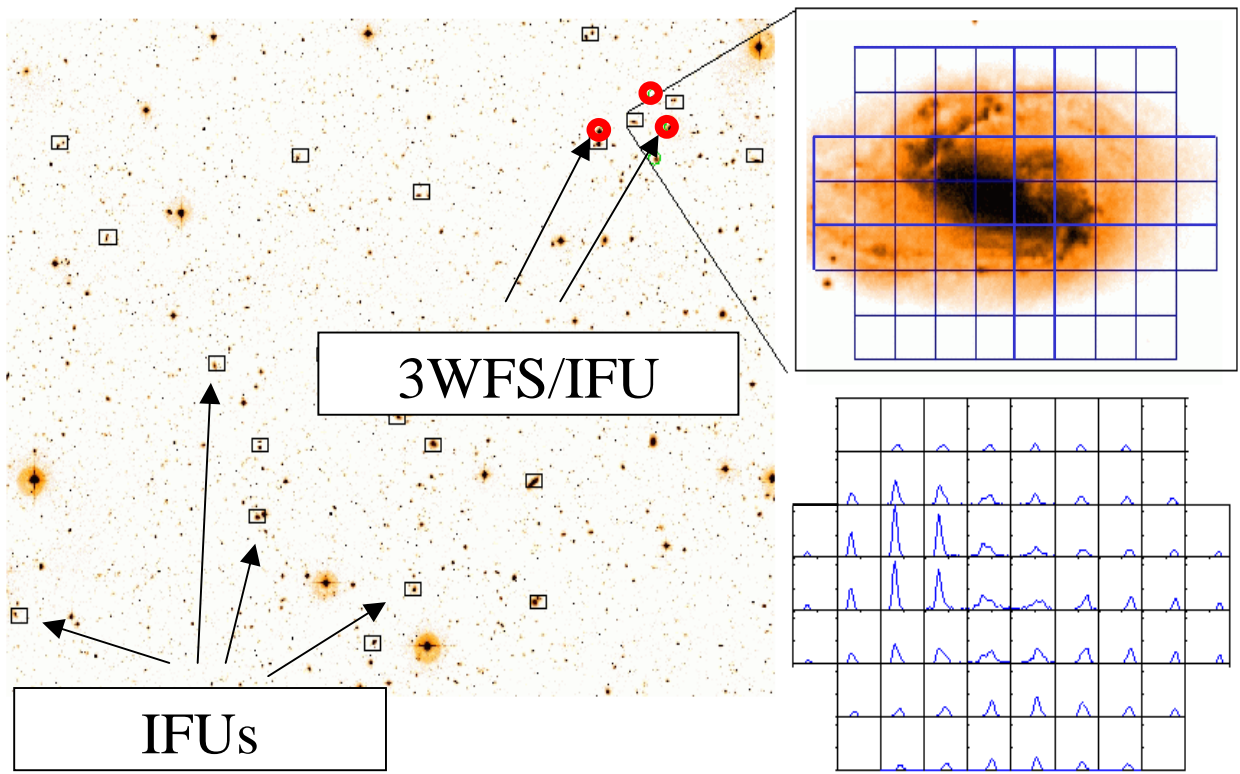

**Figure 3: a schematic view of the FALCON concept with several IFU in a wide field (Φ=25 arcmin). Each IFU integrates an AO button for correction and is coupled with 3 WFS buttons which provide a WF reference at 3 nearby R<16 stars. Each individual IFU provides 62 spectra per galaxy at a spatial resolution of ~ 0.25 arcsec (sampling ~ 0.1 arcsec).**

So miniaturization of components allows to use several AO systems at the same time, and we have to investigate the sky-coverage issue. Moreover, it has been shown in previous section that classical AO, LGS or MCAO are not optimal for FALCON. We propose to use atmospheric tomography[25,26,27] to solve the sky coverage problem. By using several WFS measuring the wavefront from 3 off-axis GS located around the IFU which samples the galaxy, one can estimate the wavefront coming from this galaxy, deduce an optimal correction in the direction of the scientific object, and then control the micro-DM of the adaptive button (see Figure 3). In our case, we are going to use "*Multiple Atmospheric Tomography*" as the process of on-axis wavefront reconstruction from off-axis measurements will be repeated as many times as there are spectroscopic IFU.

On-axis wavefront reconstruction from off-axis measurements can be considered as a linear problem. The key-point is the Reconstruction Matrix R which gives the expression of the on-axis wavefront $\phi_{gal}$ from off-axis measurements $\phi_{mes}$. On the Zernike polynomials basis[28], we have derived an expression of the optimal reconstruction matrix which minimizes the variance of the residual wavefront[29], $<\|\phi_{gal} - R.\phi_{mes}\|^2>$. It requires some knowledge of the turbulence distribution (altitude of the turbulent layers and strength of the turbulence in each layer) and the measurement noise variance on each off-axis GS. This matrix R is the product of two terms : a projection matrix T which sums the contribution of the different turbulent layers for the on-axis target (the galaxy) and a tomographic matrix W which gives the expression of the phase in the volume from the off-axis measurements. The constraint is that the use of turbulence knowledge requires to work in open-loop, whereas some recent studies have been made on its use in closed-loop mode[30].

## 4. EXPECTED PERFORMANCES AND SIMULATIONS

We have developed a simulation tool which computes an AO corrected PSF at different wavelengths with the tomographic method[29]. We have assumed a 8-meter telescope and a seeing of 0.81 arcsec at 0.5 μm (median seeing at Paranal), leading to $r_0$ (the Fried parameter[31]) of 12.7 cm. The turbulence profile includes 3 layers at altitudes of 0, 1 and 10 km with respectively 20%, 65% and 15% of the total turbulence, leading to an isoplanatic angle $\theta_0$ of 2.42 arcsec (median isoplanatic angle at Paranal) at zenith. All the turbulent layers were simulated by Fourier-filtering methods. A sample of 100 GS triplets with R < 16.5 coming from a cosmological field at a high galactic latitude (b ~ 90°) was used. For each configuration, on-axis AO corrected PSF in J and H bands were computed. Correction with Zernike polynomials went from radial order $1 < n < 14$ ($2 < j_{max} < 120$). We assumed to be in a regime dominated by photon noise. In that case the noise variance is proportional to $(N_{ph}^{-2})$ [17], where $N_{ph}$ is the number of photoelectrons per frame. Shack-Hartmann (SH) WFS were considered, leading to a propagated noise variance on Zernike polynomials following a law in $(n+1)^{-2}$ [17]. In the following simulations, the limiting magnitude was the one for which the noise variance was equal to the turbulent variance of the angle of arrival for a subaperture (~ 250 rad$^2$). Only spatial aspect of phase reconstruction was studied, and no temporal error was introduced. For each PSF, the fraction of light entering in a square aperture of 0.25 x 0.25 arcsec$^2$ was computed in J and H bands.

Our goal is to have a probability of 50% (median case, 50% of sky coverage) to gain at least a factor of 2 between the fraction of light without and with AO correction. We can see on those figures that this performance is reachable. As an example, figure 4 shows that for J band, the median fraction of light is 30 % if the first 70 Zernike polynomials were corrected compared to 15 % without correction. In H band, this performance can be expected more easily, with the correction of the first 45 Zernike polynomials which leads to a fraction of light of 35 % compared to 17.5% without correction. In the case of a higher limiting magnitude (R=17, see figure 5), this factor of 2 can be obtained with a sky coverage of 50% with the correction of the first 45 Zernike polynomials in J band and the first 35 Zernike polynomials in H band, which is understandable as in that case, noise variance becomes smaller for R < 16.5 stars. This shows that atmospheric tomography could provide a huge gain in sky coverage (from 1% to 50%).

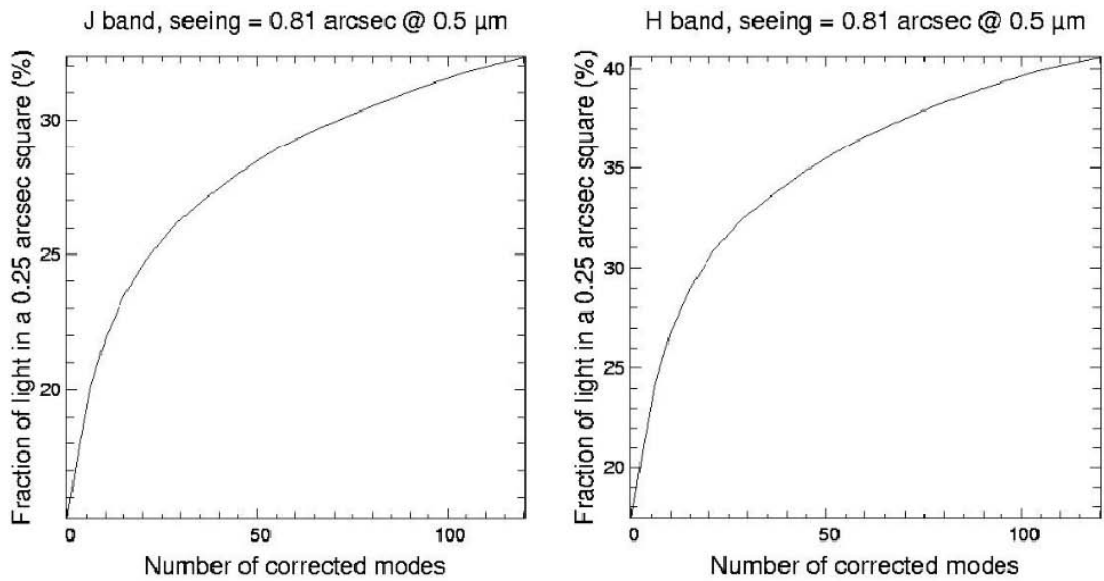

**Figure 4 : Median fraction of light entering into a 0.25x0.25 arcsec² square aperture plotted against the number of corrected Zernike polynomials. Limiting magnitude is R=16.**

As said previously, only spatial aspect was considered and no temporal error was introduced in those studies. So those results are maybe a little optimistic, as the addition of a temporal error could lead to an increase of the residual variance. We assume that the results presented on figure 4 are equivalent to the case of a R=17 limiting magnitude coupled with a temporal error. Temporal aspect will be the object of further studies.

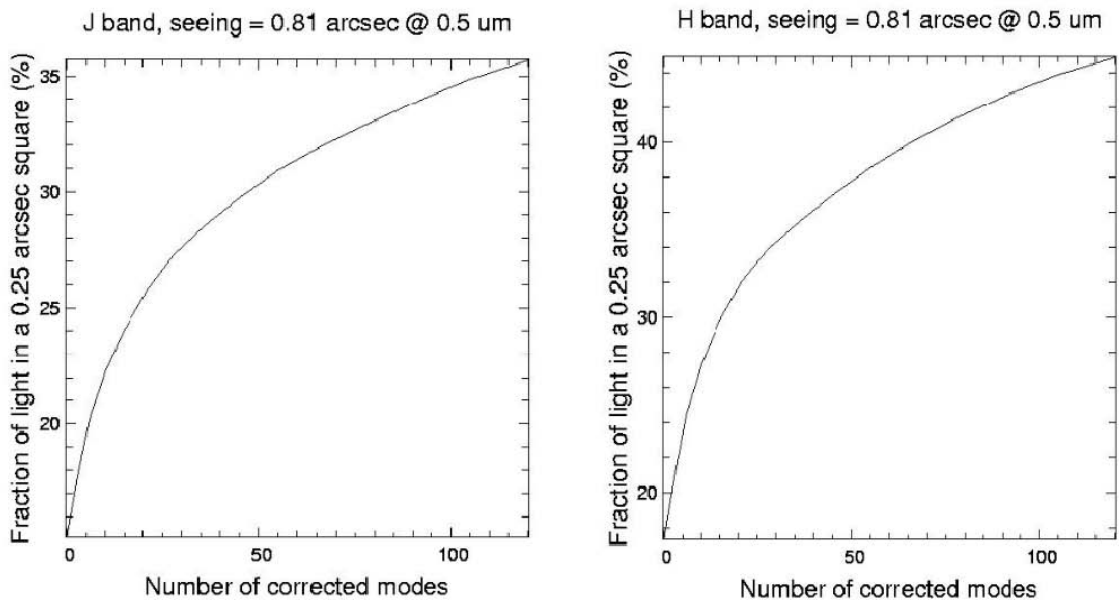

**Figure 5 : Median fraction of light entering into a 0.25x0.25 arcsec² square aperture plotted against the number of corrected Zernike polynomials. Limiting magnitude is R=17.**

FALCON at a galactic pole field can provide, for 50% of the sources, a gain in light concentration by a factor >2 if compared to natural seeing conditions. Given the fact that cosmological fields are between b=45° to b=90°, we believe that the number of requested actuators might be lowered to ~ 35-45 in order to get similar gains in using tomography techniques with at least 3 very sensitive WFS. More simulations are required to estimate the optimal micro-DM for FALCON.

## 5. INSTRUMENTAL ASPECTS

As pointed out in section 1, multi-object 3D integral field spectroscopy is the best way to study cosmological fields, allowing to reach both spatial and spectral high resolution. Previous section has shown how to overcome the spatial resolution limitation due to the seeing thanks to a new concept of adaptive optics entirely dedicated to cosmological fields. Classically, the highest the spatial resolution, the worst the spectral SNR, since there is less light entering a micro-lens of the IFU. With "Multiple Atmospheric Tomography" as we plan to use in FALCON, this constraint is overcame since light can be re-concentrated in the micro-lens by a factor ~2.

Now, we can examine the assumption made in section 3 concerning the miniaturization of the AO systems (see also Puech & al, 2003[32]). This is a crucial point since the button size (WFS and IFU buttons) limits the sky coverage in the focal plane of the instrument: on the VLT, a 50 mm diameter IFU prohibits the access to GS separated by less than 1.5 arcmin, which will restrict the number of accessible GS. So care must be taken on the IFU size, and miniaturized devices are under development.

To demonstrate the FALCON principle, we plan to use an OKO micro-DM[33]. As other DM used in classical AO, it seems not well suited to correct the tip-tilt[34] and the whole defocus modes, because of the higher dynamics required for the low order modes. Several solutions have been considered for tip-tilt correction: the micro-DM can be used on a tip-tilt mount or a micro tip-tilt mirror or an adaptive lens can be used in addition to the micro-DM.
We think that the first two solutions are not viable: the first one because of a possible excitation of an eigen mode micro-DM's membrane due to the vibration caused by the tip-tilt mount, and the second one increases the IFU size and does not allow to correct the defocus mode. So we have first preferred to evaluate the capabilities of the adaptive lens, because it is a refractive component (it doesn't need an additional reflection of the light beam, although a special care has to be brought to correct lens's chromatic effects) and it can correct tip-tilt and defocus modes, which results in a gain of dynamics for higher order modes correction by the micro-DM.
Hardy & Wallner[35] have experimentally demonstrated the possibility to use the five degree of freedom of adaptive lenses to correct low-order zernike modes. As the astigmatism modes correction requires several degrees of rotation and introduces additional aberrations, we have preferred to avoid the astigmatism modes correction by the lens. We have performed some optical simulations with a total seeing of 0.6 arcsec, in order to validate the lens corrector in the FALCON framework. Figure 6(a) represents a point array image through the system composed by the atmosphere, the VLT and a FALCON IFU-type lens corrector. The simulation comprises tip-tilt and defocus at their maximal statistical acceptable values (each mode is taken at a value equal to their 3 sigma threshold calculated in the Kolmogorov atmospherical turbulent model). These simulations show what could be optically expected through an IFU during a "short" exposure (the field is shown "frozen" on the maximal shift due to tip-tilt image motion). If a correction is applied to the lens, one can see on Figure 6(b) the resulting point array image that have to be compared with the Figure 6(c) (without atmospherical aberrations): a preliminary budget error has shown that the optical resolution achieved by this corrector can be near of 0.3 arcsec at the IFU field center[36]. We underline that this simulation does not take into account the micro-DM and that the values taken for each mode are very pessimistic, but if the system can operate a good correction in this configuration as it seems to be, it will be efficient too in a more favorable situation.

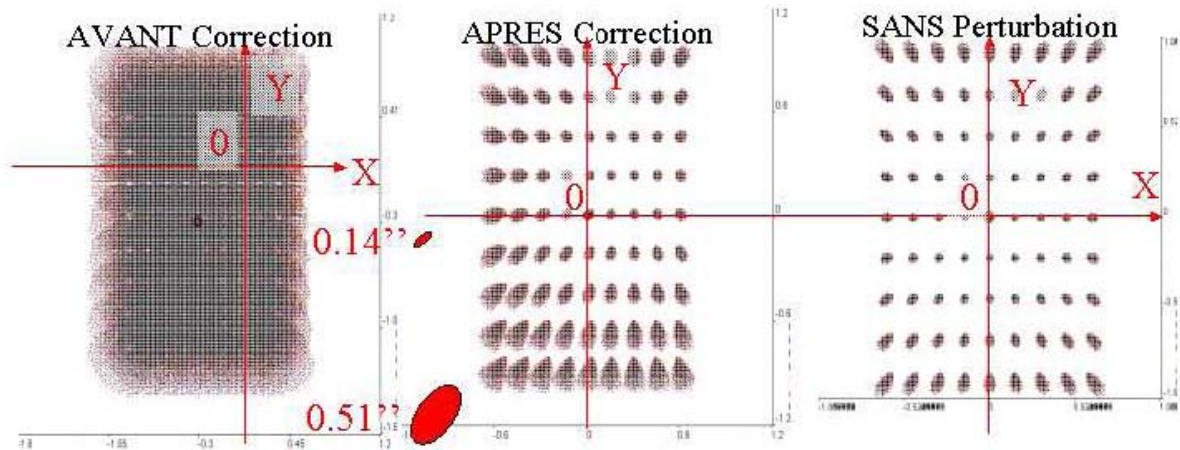

**Figure 6 : Simulations of the lens corrector used in FALCON. (a) [left] image of a grid through the turbulent atmosphere (see text).(b) [middle] the same after correction by the lens.(c) [right] without turbulence case.**

Some studies are currently made at Observatoire de Paris about the optomechanical design of an adaptive button with an adaptive lens, and a OKO micro-DM. First results with present technologies show that it should be possible to build a device of 50 ×50 ×200 mm including all those components. A preliminary design can be seen in Figure 7, where light enters the button on the right side. As one can see on this figure, the adaptive lens is actuated by piezo-stack elements. The OKO mirror is on the left and the light exit the button through optical fibers (not represented here) at the bottom. Such a design is preliminary and several alternatives are also studied.

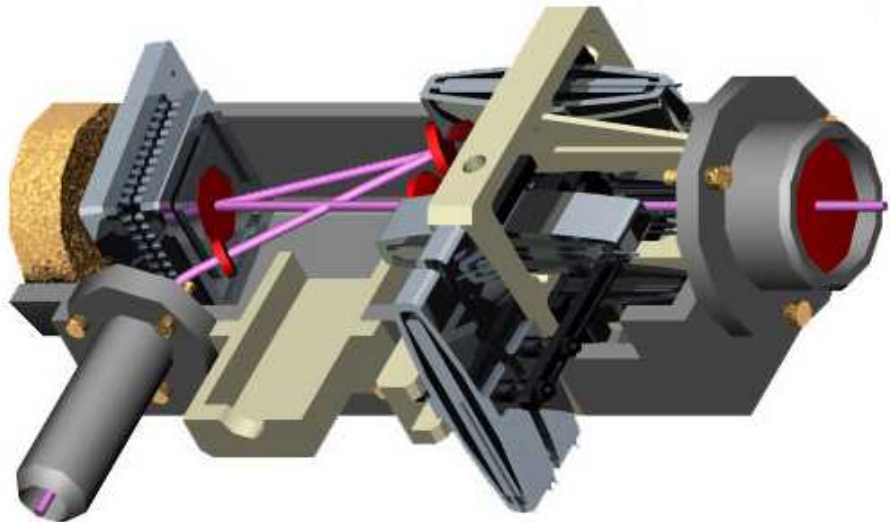

**Figure 7 : preliminary opto-mechanical design of the FALCON IFU button.**

The spatial scale (in arcsec by mm) in the focal plane of a telescope grows with its focal length or, at constant apperture, with the telescope diameter. That is why ELT can relax the miniaturization constraint on the buttons width: FALCON should be easier to built on ELT than on 8-meter telescope. Another advantage is the effect of the outer scale $L_0$ which reduces amplitudes of correction when the telescope diameter is no more negligeable compared with $L_0$ [37]. However, some disadvantages appear as the number of actuators required for the DM has to be higher. Specific studies will have to be done to transpose FALCON on ELT.

## 6. CONCLUSIONS

Our study of the FALCON concept is supported by contracts with the Paris Observatory, CNRS and ESO. Our study is firstly focusing at the performances of such a system if it was implemented at VLT. Our simulations show that we can reach spatial resolution lower than 0.25 arcsec, in J and H, for ~50% of sources selected in cosmological fields. The proposed concept is to correct only small areas of few arcsec$^2$ (distant galaxy size), distributed in a wide field of view, assuming specific adaptive optics systems (buttons) located at or near each of the targeted galaxies. By coupling these AO systems to a spectrograph with multiple IFUs, one can improve by a factor 2 the light concentration within an aperture of 0.25 arcsec, when compared to the natural seeing conditions. The proposed AO system requires the coupling of micro-DM with an adaptive lens which will account for the low order correction modes (tip tilt and defocus), relaxing the dynamical performances required for micro-DM. The above performances can be reached using micro-DM with ~35 to 50 actuators.

If implemented at VLT, this system has to include very small devices to avoid field obstruction. Such devices require a new generation of micro-DM and WFS. Another critical issue is the need to close or partially close the AO loop. It could be either closed by electro-mechanics similarities (between the different buttons) or by software throughout a detailed modeling of the expected wavefront at the scientific target. The system might also be coupled with a DM secondary mirror of the telescope since latter may have some limitations in correcting large fields of view.

The FALCON concept has to be extrapolated to extremely large telescopes. Indeed cosmological studies in the very distant Universe require reasonably large fields of view to study galaxy formation at scales beyond the galaxy correlation typical lengths. Simple extrapolations suggest that FALCON would make possible to detail galaxy physics down to 400 pc scales at z=2-6, and then to describe in details the mechanisms of the formation of each present day galaxy type. Miniaturization of the FALCON devices might be somewhat relaxed, although to reach such exquisite spatial resolutions would require a significantly larger number of actuators per deformable mirror. We believe that ELT equiped with FALCON would be unbeatable to understand how galaxies were formed since the epoch of reionization and beyond.